\documentclass[11pt,twoside]{article}  
\usepackage{asp2006}
\usepackage{adassconf}

\begin{document}   

%
%

\paperID{O1.1}

%

\title{Talking Amongst Ourselves --- Communication in the Astronomical Software Community}

%
%
%
%
%

\markboth{SHORTRIDGE}{Communication in the Astronomical Software Community}

%
%
%
%

\author{Keith Shortridge}
\affil{Anglo-Australian Observatory}

%

\contact{Keith Shortridge}
\email{ks@aaoepp.aao.gov.au}

%
%
%

\paindex{Shortridge, K.}

%



\begin{abstract}          
Meetings such as ADASS demonstrate that there is an enthusiasm for communication within the astronomical software community. However, the amount of information and experience that can flow around in the course of one, relatively short, meeting is really quite limited. Ideally, these meetings should be just a part of a much greater, continuous exchange of knowledge. In practice, with some notable --- but often short-lived --- exceptions, we generally fall short of that ideal. Keeping track of what is being used, where, and how successfully, can be a challenge.
A variety of new technologies such as  those roughly classed as 'Web 2.0' are now available, and getting information to flow ought to be getting simpler, but somehow it seems harder to find the time to keep that information current. This paper looks at some of the ways we communicate, used to communicate, have failed to communicate, no longer communicate, and perhaps could communicate better. It is presented in the hope of stimulating additional discussion --- and possibly even a little action --- aimed at improving the current situation.
\end{abstract}

%
%

\section{Preface}

What follows is essentially a verbatim transcript of the introductory talk as given at ADASS, in Quebec, on the morning of the first session --- which happened to be the day before the 2008 U.S. presidential elections.

\section{Introduction}
Good morning everybody. Bonjour, tout le monde. Everyone awake? No jet lag? It's very nearly tomorrow morning in Australia, which is confusing for me, so if I drift off topic, just talk amongst yourselves.

Because that's what this is all about. Talking amongst ourselves. I want to try to encourage some discussion about whether we can communicate better amongst ourselves. This seems particularly worthwhile in the context of an ADASS meeting, because this is one of the few occasions we do all get to meet together.

In many ways, {\em you} don't need to hear this talk --- but don't try to leave, there are webcams on all the doors and {\em we} know your e-mail addresses. You don't need it because you {\em are} the communicators in our field. You've come to ADASS. You've probably struggled with your organisation's bureaucracy and expenses software. (Have you noticed that lots of places seem to have accounting software that can't possibly have been used anywhere else first --- except possibly Enron and Lehman Brothers?) You've booked the tickets, carried on your laptops --- how many people wrote their talks on the way over? Oh, all still working on them, are you\ldots

You've explained to your colleagues that it's not just a jolly. It's cold in Quebec in winter. You're prepared to miss what's going on back home. Tomorrow's a big day in some places. November 4th. In Australia, they're running the Melbourne Cup, the biggest race of the year. And this year, I'm not going to lose money on it for the first time, because I'm here. And tomorrow, something important is going on south of the border, right?

ADASS is just about the main communications channel in astronomical software. The people who run it deserve some applause. But ADASS on its own isn't enough.

We should mention the SPIE meetings. SPIE is covering software now, especially at what I think of as the sharp end --- telescopes and instrument control. And this year, SPIE had a very interesting session on software sharing. I'll come back to that later.

\section{Keeping current}

Outside the conferences, and outside individual projects, things are much more limited.  And conferences don't cover everything --- they naturally concentrate on the cutting edge work. They're expensive to get to, and what people learn at them can be quite hard to pass on to others. They probably aren't going to get cheaper --- `peak air travel' and `peak oil' are probably going to happen at about the same time.

Now, I'm here because a few months ago I scattered around an e-mail bemoaning the state of communications in astronomical software. And this happened because I suddenly realised how little I knew about what was going on.

I was asked to prepare an estimate for a proposed Antarctic telescope. Produce a software structure design for the whole system --- telescope, instruments, data reduction, communications, everything. Costs accurate to ten percent would be nice, and you can do it in about two weeks, can't you?

Suddenly, I needed to know how you would structure the software for a new telescope.

I learned something a long time ago as a physics student doing lab experiments. If you're having trouble with something, find out who did it the week before and ask them. Back then, this was called `cheating'. Now we call it `consulting the knowledge base'. 

(I've always assumed this was the real driver behind projects like SETI: you want to know about the rest of the Universe? Find someone who lives there and ask them. `Excuse me, you know of any Earth-like planets in your neighbourhood?'. Much easier than trying to find them yourself, and you don't need all those expensive telescopes.)

So, I wanted to consult the knowledge base for our field. And I wasn't sure where it was.

Incidentally, at the other end of the food chain, the astronomers do this quite well. (I'm not sure just which end they are, but it's the other end.) Software people don't use libraries much, because it always seems they only have the out of date books: `8086 architecture', `VMS programming', `Field guide to dinosaurs', etc.. But astronomers are in there reading the Annual Review of Astronomy and Astrophysics, where pundits publish review papers --- quite a big job, but they get cited a lot. On the web, pre-prints appear in astro-ph. Conferences have review talks.

There isn't an Annual Review of Astronomical Software, so how do you find out what people are using? What middleware? ACS, CORBA, .net, DRAMA, TAROS? Can you really run everything from LabView? How do you string together a data reduction pipeline?

I found myself wanting some central web site with all this arranged for me. With a search box where I could type `observatory middleware'. Some if this stuff is available, but not all, and Google is only a rather awkward window into it.

Information needs an index and a table of contents, and producing really good ones is still a job for people.

In the past, there have been Web sites that have tried to provide what I, at least, can't find now --- an overview of our whole field. Some of you have been directly involved in them.

ASDS, the Astronomical Software and Documentation Service, started with the idea of helping you find astronomical software packages, and expanded to cover telescope and instrument manuals. But I'm afraid that after years of useful service, ASDS is now just a set of links to a machine that no longer responds. One of these links is in AstroWeb, another site that seems to be on the backburner now.

\section{What doesn't help}

Obviously, there are reasons why sites are allowed to fade away. Things have changed in our world over the years. Some changes are good for communication --- the way the Web itself has changed is one of these. Collaborative tools have emerged: TWikis, blogs, even YouTube and SecondLife --- I'll get back to these. But there also has to be a will to use them, and things have changed there too.

A lot of us are more `managed' these days. Time spent has to be justified and accounted for. And you aren't supposed to spend your time giving away what your organisation perceives as its competitive advantage.

To me `IP' stands for `Internet Protocol' and is a good thing. Unfortunately, it also means `Intellectual Property', and that may or may not be a good thing. OK, I can see why this matters. Organisations have to pay the rent: `Spare a dollar? Telescope and ageing instrument suite to support', `Every cent donated goes to our latest design study'.

And security is an issue. I actually found it hard to get all the e-mail addresses I wanted, even for people I knew, because many places don't publicise e-mail addresses any more. Because if your address gets out, you start to get such interesting e-mail. Someone in Singapore regularly sends me a list of prices for second-hand road-building equipment. `Caterpillar excavator, good condition, engine recently rebuilt.' How they found out I need one is beyond me.

None of this helps. It is hard to find the time to maintain Web pages for the public good, even if you're part of that public, and to justify it to management.

But, at the same time, the Web is making things much easier. And I suspect most of you are already using some of these new tools --- at least within specific projects.

\section{What does}

You can see why public interest, general purpose, Web pages fade. If all changes have to go through one or two central people, making these changes will become dull, burdensome, unprofitable. But this is the new millennium. 20th Century Fox may have decided to stay in the last century (and a lot of marketing dollars must have gone into that decision) but the Web has moved on.

The point about all the `Web 2.0' technologies is that their content is no longer centrally provided. Wikipedia, YouTube, Flickr, are just repositories for user-created content.

Lots of you use TWikis for your collaborative projects. People hold video conferences over Skype. Way back when international phone conferences were a very big deal, I once sat in on a meeting in the UK from my office in Australia using e-mail. I had a contact at the meeting, and when they wanted my opinion, a couple of e-mails went back and forth. It actually worked pretty well, given that it was late afternoon in the UK and sometime after midnight in Australia. And then, around three-thirty in the morning, I realised I'd not heard anything for quite some time, and nobody was replying to me any more. I waited some time, then gave up and went home. Later, I discovered they'd finished the meeting, packed up, and gone to the pub for a drink. Not only didn't they invite me, they didn't even bother to tell me\ldots

Skype is {\em much} better. And people really do hold virtual conferences in SecondLife, and presumably their avatars all go off to a virtual pub afterwards and get virtually drunk. And it's these extra meetings, of course, where you learn a lot at conferences, but which never gets into the proceedings.

In various ways, individual projects are, individually, making good use of the Web 2 tools. Some of them. One point made more than once in the replies to my grumbling e-mail was that the VO people did a better job of communicating than do the people at my end of things --- the instrument and telescope control people.

I suspect, although it might be a topic for discussion later, that VO is set up as a really big collaborative project, and that in such projects internal communication is expected, budgeted for, supported, has a Microsoft Project task code, and so on. So even writing a project blog is a legitimate activity, as it should be.

But outside individual projects, if you want to see the whole range of activity across our field, what is there? Not a lot.

\section{New initiatives}

When I sent my e-mail around, I didn't know anything about the planned software sharing initiative at SPIE this year. A paper on software sharing was presented, and an open discussion followed. Some of the people involved are here --- Alan Bridger, Kim Gillies, Steve Wampler, should all be in this room somewhere.

You can get a copy of the paper `Enabling Technologies and Constraints for Software Sharing in Large Astronomy Projects' (Chiozzi et al. 2008), if you know what to Google for. It shows up on the ESO Web pages, and you can get it from SPIE itself --- if you're a member, or know someone who is.

It's an interesting paper. It identifies a lot of what makes software sharing difficult. I was interested to see that they quote the example of the portable telescope control system, PTCS. They comment that it has been adopted by a number of telescopes, but `we could find no information about recent developments'. And, as I read that, we were commissioning the replacement control system for the AAT, the Australian 4 meter, completely based on PTCS. So there certainly had been recent developments. It's an example of poor communication, I guess. I don't think PTCS has a public Web presence.

I agree it's hard to share software --- although, obviously, people do. But I think it's much easier to share experiences. Even if you could find PTCS for download on the web, what you {\em really} want first would be to see comments from a number of people who've done telescope control systems, explaining what they did and how well they thought it worked.

As a result of this software sharing initiative, there is, of course, now a TWiki\footnote{www.astroshare.org}. Right now, it's brand new, fairly empty. Getting content won't happen by magic. I suspect it will need people to solicit content --- I heard a suggestion for a rotating `editor' position, for example. 

Here's a idea. If someone from each institution here just listed the software they used --- instrument control environments, middleware, data reduction pipelines --- even their project management software --- that would be a lot of organised information that could go in a TWiki.

I'd like to ride a personal hobby horse here. Documentation doesn't have to be beautiful to be useful. Some things --- geometric algorithms, say --- need complex diagrams, and Microsoft Word isn't very good at doing them. They take ages to do on a computer. Or a few minutes on a white board or a piece of paper, then a digital camera or a scanner and you have a JPEG diagram\footnote{In this spirit, this talk was originally written by hand on sheets of paper, and a computer-readable backup copy was made by photographing them. These were then used to produce this paper.}. Video yourself drawing it on the white board, and you can explain the algorithm as you go.

Our one AAO expert on the VLT instrumentation software left us, and during his last week I sat him down with a white board and a terminal, in front of a video camera, and got him to explain the system internals to me. I'd ask him about what I didn't understand, he'd explain the short cuts that usually don't get documented. I learned a lot --- I learned you should turn your mobile phone off before doing this stuff --- and we now have some permanent documentation.

You can put your documentation on YouTube if you like. You can certainly put videos into a TWiki.

\section{Finally}

One reply to my original e-mail put it bluntly: `You have to answer the question: What's in it for me?' Well, apart from the fame, the glory, the adulation, the hundreds of screaming fans\ldots Oh, sorry, that's a different career path.

But you know the answer. It's why you really are the right audience for this talk, and I'm glad you stayed for it. You know what you get from communication, from sharing experiences. It's why you've come all this way with your laptops and no more than 100 ml of liquids in your hand baggage.

It's because it makes you job more interesting, easier, more satisfying. It's because a lot of people building on each other's work can build something bigger and better.

And now you've had half an hour of me telling you what you really knew anyway, we can get down to the real work of the conference. I'll see you all in a bar somewhere\ldots

\end{document}